\DocumentMetadata{}
\documentclass[hyphens,nonacm,sigplan]{acmart}

\usepackage{amsfonts}
\usepackage{algorithm}
\usepackage{algorithmic}
\usepackage{graphicx}
\usepackage{textcomp}
\usepackage{xcolor}
\usepackage{multirow}
\usepackage{multicol}

\usepackage{listings}

\definecolor{codegreen}{rgb}{0,0.6,0}
\definecolor{codegray}{rgb}{0.5,0.5,0.5}
\definecolor{codepurple}{rgb}{0.58,0,0.82}
\definecolor{backcolour}{rgb}{0.95,0.95,0.92}
\lstdefinestyle{mystyle}{
    language=c++,
    backgroundcolor=\color{backcolour},
    commentstyle=\color{codegreen},
    keywordstyle=\color{magenta},
    numberstyle=\tiny\color{codegray},
    stringstyle=\color{codepurple},
    basicstyle=\ttfamily\footnotesize,
    breakatwhitespace=false,
    breaklines=true,
    captionpos=b,
    keepspaces=true,
    numbers=left,
    numbersep=5pt,
    showspaces=false,
    showstringspaces=false,
    showtabs=false,
    tabsize=2
}

\lstdefinestyle{verilogStyle}{
    language=Verilog,                               
    basicstyle=\footnotesize\ttfamily,               
    keywordstyle=\color{blue},                       
    commentstyle=\color{green!50!black},             
    stringstyle=\color{red},                         
    showstringspaces=false,                          
    numbers=left,                                    
    numberstyle=\tiny\color{gray},                   
    breaklines=true,                                 
    breakatwhitespace=true,                          
    tabsize=2,                                       
    frame=single,                                    
    captionpos=b                                     
}

\lstdefinestyle{cppstyle}{
    language=C++,
    basicstyle=\ttfamily\small,
    keywordstyle=\color{blue},
    commentstyle=\color{green!60!black},
    stringstyle=\color{orange},
    numbers=left,
    numberstyle=\tiny\color{gray},
    stepnumber=1,
    showstringspaces=false,
    breaklines=true,
    breakatwhitespace=true,
    tabsize=2,
    frame=single,
    captionpos=b,
    belowcaptionskip=1em,
    mathescape=true,
}

\usepackage[]{hyperref}

\begin{document}

\title{Aligning Netlist to Source Code using SynAlign}

\author{Sakshi Garg}
\email{sgarg3@ucsc.edu}
\affiliation{%
Department of Computer Science and Engineering\\
  \institution{University of California, Santa Cruz}
  \city{Santa Cruz}
  \state{California}
  \country{USA}
}

\author{Jose Renau}
\email{renau@ucsc.edu}
\affiliation{%
Department of Computer Science and Engineering\\
  \institution{University of California, Santa Cruz}
  \city{Santa Cruz}
  \state{California}
  \country{USA}
}

\begin{abstract}
    In current chip design processes, using multiple tools to obtain
a gate-level netlist often results in the loss of source code correlation.
SynAlign addresses this challenge by automating the alignment process,
simplifying iterative design, reducing overhead, and maintaining correlation across
various tools. This enhances the efficiency and effectiveness of chip design workflows.

Improving characteristics such as frequency through iterative design
is essential for enhancing accelerators and chip designs. While synthesis
tools produce netlists with critical path information, designers often lack the
tools to trace these netlist cells back to their original source code. Mapping netlist
components to source code provides early feedback on timing and power for
frontend designers.

SynAlign automatically aligns post-optimized netlists with the original source
code without altering compilers or synthesis processes. Its alignment strategy
relies on the consistent design structure throughout the chip design cycle, even
with changes in compiler flow. This consistency allows engineers to maintain
a correlation between modified designs and the original source code across various tools.
Remarkably, SynAlign can tolerate up to 61\% design net changes without impacting
alignment accuracy.
\end{abstract}

\maketitle
\pagestyle{plain}

\section{Introduction}\label{sec:intro}

\begin{figure}[htpb]
  \centering
    \includegraphics[width=\linewidth]{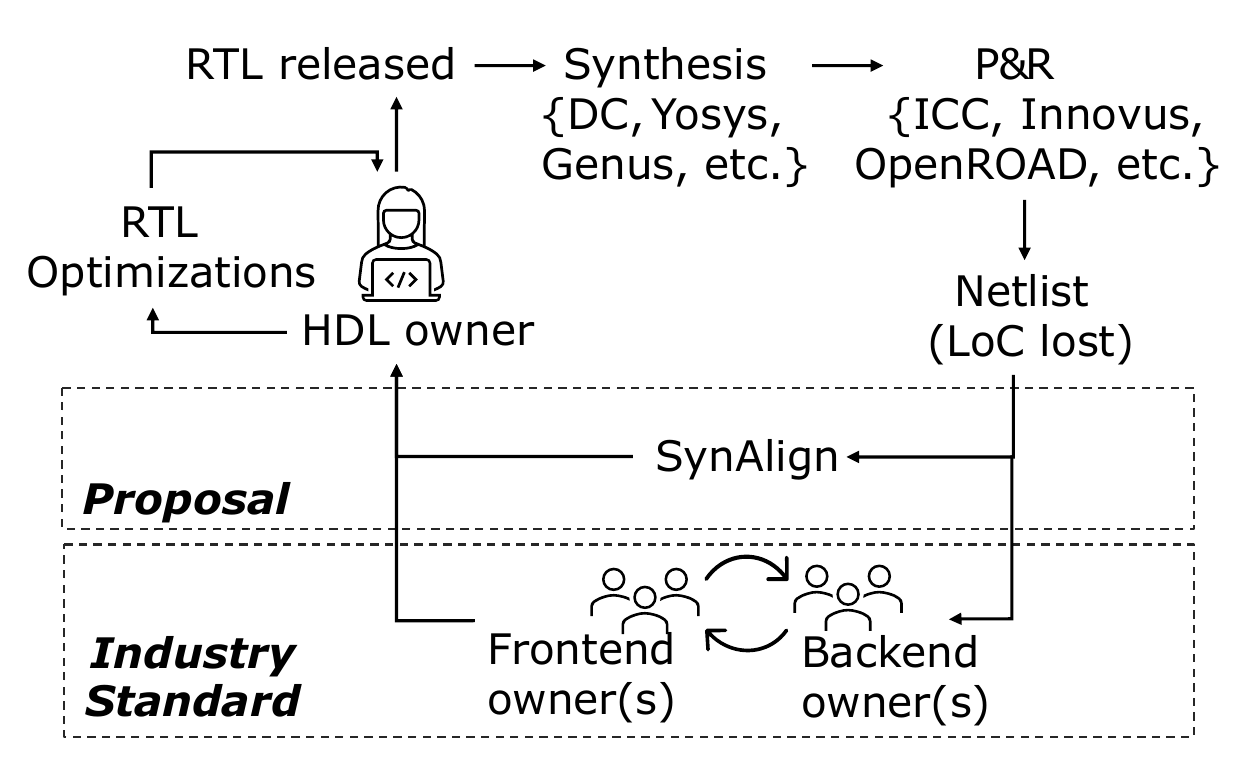}
  \caption{SynAlign Vs. current industrial practice.}
  \label{fig:proposal}
\end{figure}

Chip design requires multiple iterations of microarchitecture changes
to improve performance metrics. These iterations involve multiple tools that
perform synthesis,
flattening\footnote{Flattening resembles function inlining in non-hardware
flows.}, placement, and routing. 
Many tools further modify the netlists, resulting in significant differences from the original source code.
Each iteration also necessitates efforts to
correlate parts of the netlist to the source code to maintain alignment.
Tools like PrimeTime~\cite{synopsys:primetime}
and OpenSTA~\cite{opensta} provide netlist
insights, but designers need to manually trace net names back to
their source code locations, which is complex and time-consuming. Our paper aims
to automate this alignment process as shown in Figure~\ref{fig:proposal}.

Hardware flows, unlike software compilers, involve interprocedural
optimizations and multiple tools within a single chip design cycle.
Software compilers like LLVM use ``Source Locators'' as
metadata~\cite{8639205, clang-locations}, requiring consistent
propagation of source locator information across all code transformations.
While some FPGA tools also propagate source location embeddings, their limited
scope and restriction to a single tool reduce the choice of compilers for chip
design. This limitation is particularly challenging due to the lack of source
locator sharing between different tools and the extensive modifications
introduced by synthesis.

Implementing source locators is
resource-intensive~\cite{LLVMdocumentation,
Update_Debug_Info_LLVM} and can introduce backward
compatibility issues~\cite{chisel2022}.
Significant structural changes, abstraction level shifts, and optimizations
during synthesis complicate maintaining a direct correspondence between the
original HDL code and the synthesized hardware. Transformations such as resource
sharing, technology mapping, and inlining often result in hardware that no
longer resembles the original RTL description, making it difficult to accurately
track the original line of code through these transformations.

The effort required to implement source locators is exemplified by the Chisel
team, which began embedding source information in 2016 and has continued
updating it across various
releases~\cite{chisel2016,v355chisel2001,vM2,vM1,v350chisel1}. This
process is resource-intensive~\cite{LLVMdocumentation,
Update_Debug_Info_LLVM} and can complicate code, introducing backward
compatibility issues~\cite{chisel2022}.

The current chip design process typically involves weekly meetings between
backend and frontend teams to discuss timing information and map critical paths.
Due to the complexity of this task, some teams avoid aggressive optimization
techniques like flattening to simplify the process.

We propose SynAlign, a tool that automatically maps annotated post-optimized
netlists to the original source code. SynAlign leverages structural equivalence
points to align nets in the netlist with their source code counterparts. By
avoiding the need for source locators, SynAlign reduces compiler design
overhead and eliminates the costs associated with source metadata, simplifying
the process for compiler developers.
Furthermore, SynAlign allows engineers to maintain a correlation between
modulated designs and source code across multiple tools and Hardware Description Languages (HDLs).

Mapping remains possible without source locators because synthesis tools,
such as Yosys~\cite{yosys} and Design Compiler~\cite{synopsys:design-compiler},
attempt to preserve some net names,
particularly sequential ones like registers and memories, when feasible. This preservation aids Formal
Equivalence Check (LEC)
tools~\cite{lec_des,netolicka2005equivalence} and supports engineers. However, net names often change
to temporal names when their meaning changes, and typically, only 5\% to 20\% of
net names are preserved during synthesis.

SynAlign uses design inputs/outputs and preserved net names as Anchors. Through
an iterative process, SynAlign aligns equivalent nets between graph
representations of the source code and the transformed or synthesized code.
Notably,
SynAlign does not impose any extra limitations on optimization when using these
anchors.

To evaluate SynAlign, we tested various designs written in Chisel~\cite{chisel}
and Verilog~\cite{thomas2008verilog,palnitkar2003verilog}.Chisel designs present
a unique challenge because Chisel, a Scala DSL, must be compiled into Verilog
before synthesis. In contrast, Verilog designs are simpler to map as they do not
require this additional generation step. 

Across all examined designs, SynAlign correctly identifies the source code
location 50-93\% of the time. This represents the first automated correlation
between netlists and source code separated by multiple tools, with no existing
baseline for comparison. Our scalable approach contrasts with current solutions
that require human intervention, often taking hours or days. SynAlign operates
much faster than typical chip design iteration times. For example, when mapping
the critical timing path from netlist to source code in a design like
Rocket~\cite{Asanović:EECS-2016-17},
SynAlign achieves 77\% accuracy in less than 20 seconds.

In summary, our main contributions are:
\begin{itemize}
\item {\textbf{Automating back annotation} between \textbf{fully flattened and optimized} designs and
their hierarchical source code. Thus enabling chip designers to leverage full
optimization capabilities while quickly relating the netlist to the source code
for further iterations.}
\item {Correlating different hardware design stages
extracted from \textbf{multiple tools} within the same chip design cycle.}
\item {Providing an approach applicable to \textbf{any hardware design language (HDL)} that
can be translated to equivalent Verilog with Line-of-Code (LoC) information.}
\item {Introducing the novel use of network alignment strategies to align synthesized
netlists with their source code, exploring and utilizing hardware design
attributes for comprehensive and efficient design analysis.}
\end{itemize}

\section{SynAlign Algorithm}\label{sec:methodology}

\begin{figure*}[!ht]
  \centering
      \includegraphics[width=\linewidth]{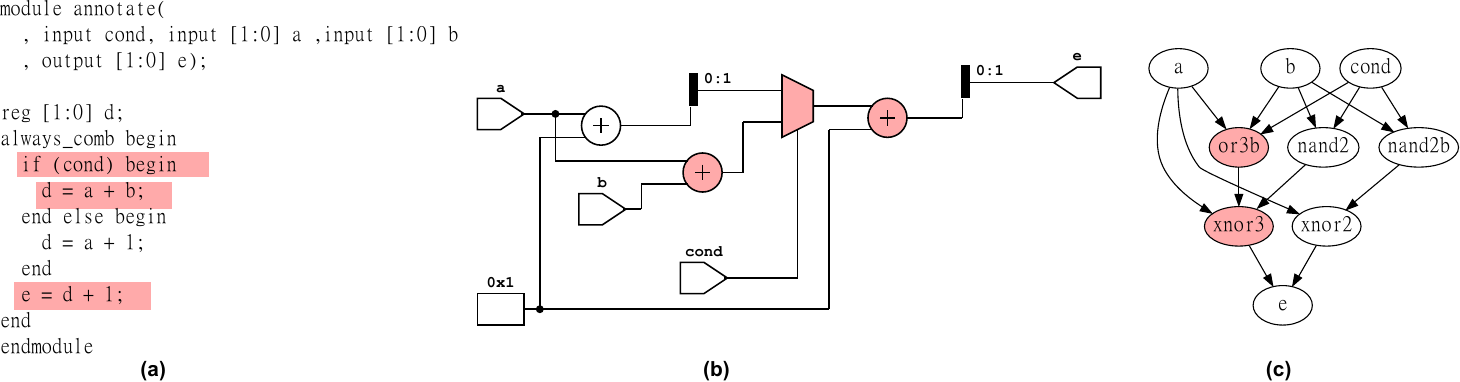}
  \caption{(a) Source code example with aligned part highlighted. (b) Reference graph ($G_{ref}$). (c) synthesized graph ($G_{synth}$) with annotations.}
  \label{fig:annotate}
\end{figure*}

SynAlign aims to identify equivalence points between two networks: the graph
representation of the HDL source code ($G_{ref}$), which includes source code
location information, and the graph of synthesized netlist ($G_{synth}$). Both
are directed graphs where nodes have multiple drivers (outputs), sinks
(inputs), and edges (or net). A user or tool annotates nets in
$G_{synth}$, and SynAlign matches them to the original source code ($G_{ref}$).

Figure~\ref{fig:annotate} illustrates a Verilog code snippet
(Figure~\ref{fig:annotate}.a), the
corresponding high-level structural representation (Figure~\ref{fig:annotate}.b) generated using
netlistsvg~\cite{netlistsvg}, and the synthesized graph ($G_{synth}$)
(Figure~\ref{fig:annotate}.c) obtained
using Yosys~\cite{yosys}, skywater130 liberty.
The critical path, highlighted in red, includes cells `or3b' and `xnor3'.
Despite the simplicity, the mapping process is intricate, aligning gates
with structural mapping operators like adder and mux. The pertinent
lines of code are highlighted in the figure.

\begin{center}
\begin{lstlisting}[style=mystyle, caption={SynAlign algorithmic overview},
label={listing:SynAlign_algo},escapechar=|]
Anchor Point Matching  (Section|~\ref{subsubsec:anchor}|)
Topological traversals, calculate pending nets
for pending sequential nodes:
  Full+Half Matching (Section|~\ref{subsection:fullHalfMatch}|)|\label{line:seqFullHalfMatch}|
  Partial Matching (Section|~\ref{subsec:partialMatch}|)
for pending combinational nodes:
  Full+Half Matching (Section|~\ref{subsection:fullHalfMatch}|)
  Surrounding Matching (Section|~\ref{subsubsec:surrCell}|)|\label{line:surrMatch}|
  Partial Matching (Section|~\ref{subsec:partialMatch}|)
Print the annotated net and its aligned net(s) with the LoC
\end{lstlisting}
\end{center}

After the annotation of nets in $G_{synth}$, the alignment process detailed in Listing~\ref{listing:SynAlign_algo},
involves identifying Anchor Points. Anchor
Points are graph nets with the same names in both $G_{ref}$ and $G_{synth}$. The
number of Anchor Points depends on the synthesis tool, but at a minimum, both
graphs share the same top-level inputs and outputs as Anchor Points. During
Synthesis, original nets may split into various nets or registers, creating a
one-to-many mapping. Using Anchor Points as initial Resolved Points (RPs),
SynAlign analyzes RPs in both the $G_{synth}$ and $G_{ref}$. Resolved Points are
the nets in both the graphs
that have been confidently aligned as equivalent. These RPs are calculated
for each graph node driver through topological traversal and include Start
Resolved Points (SRPs) and End Resolved Points (ERPs).
An RP connected
through input nets is considered an SRP, whereas an RP connected through output
nets is an ERP.

To reduce the compute costs, SynAlign employs two main loops as shown in
Listing~\ref{listing:SynAlign_algo}: first to align the
sequential subgraph and the next for the pending-to-identify combinational subgraph.
Nets aligned during
matching passes also function as RPs for subsequent stages.
Nets with identical Start and End RPs are marked as
equivalent, and this process continues until no new matches are found. 
Unresolved nets may persist due to factors like optimized-away start points. If
surrounding netlist nodes correspond to the same code line, their mapping is
expanded. Nets with the highest number of equivalent matches are also
considered for alignment. This method enables engineers to trace each net back
to its original source code, ensuring high alignment accuracy.

With these prerequisites in place, we perform the alignment process outlined in
Listing~\ref{listing:SynAlign_algo}. Each alignment stage is discussed in detail subsequently.

Before identifying the pending combinational cells, SynAlign
performs full and partial alignment only for a subgraph consisting
of sequential cells and known RPs. Inn digital systems like
RocketTile~\cite{Asanović:EECS-2016-17}, this subgraph represents only 14\% of
the entire $G_{synth}$, significantly reducing the problem size.

After aligning all sequential cells, SynAlign iterates over the pending
combinational cells to perform full and partial alignment. These steps reduce
time complexity to less than quadratic because, in pipelined designs,
combinational cells are separated by sequential cells. This makes
SynAlign more efficient than traditional NP network alignment
problems~\cite{zhang2016final}, which typically have
quadratic~\cite{zhang2016final} or cubic complexities~\cite{heimann2018regal}.

\subsection{{\bf Anchor Points Matching}}
\label{subsubsec:anchor}

Anchor Points are nets with identical hierarchies and names in both graphs,
indicating nets that have retained their names throughout the synthesis cycle,
allowing confident network alignment. The datatype of the driver node for nets
identified as Anchor Points is also the same.

Anchor point matching starts by addressing nomenclature inconsistencies
from different synthesis tools. For example, some compilers might name buses as
$var\_1\_$, while others use $var[1]$. Unlike Yosys, Design Compiler (DC)
appends $\_BAR$ to certain variables.
Hierarchy naming also varies, with some tools using ``.'' and others using ``\_''
to separate hierarchies.
Each net in both the graphs is renamed to follow same nomenclature.
This net name processing can be enhanced to accommodate various synthesis nomenclatures.
Also, synthesis tools are designed to avoid
generating new names that inadvertently match $G_{ref}$ net names, as formal
logical equivalence tools also rely on Anchor Points.

Initially, we identify Anchor Points by hierarchically traversing
$G_{ref}$, treating these points as RPs for subsequent matching references.
During this stage, $G_{ref}$ is traversed first to capture connections as hashmaps
for correlation during $G_{synth}$
traversal. Assuming $m$ edges in both the graphs, the total time complexity of this stage
comes out to be $O(m) + O(m) = 2 * O(m)$.
The cost of topological traversal is amortized by
computing it once beforehand. A pending-points map is also created
for each net in $G_{synth}$. Anchor Points are then omitted from the pending-points
map to reduce the overall algorithm's time complexity.

\subsection{{\bf Full+Half Matching}} \label{subsection:fullHalfMatch}

Full match occurs when RPs in $G_{synth}$ match all RPs in $G_{ref}$.
If $G_{synth}$ has a net name $N_s$ and $G_{ref}$ has a net name $N_r$,
and both their SRPs and ERPs are identical, then Nr \textit{fully matches} Ns.
This principle underlies the concept of \textbf{Full Matching}.

Alignment is then performed
for the ``next most matching net name''. If either SRPs or ERPs are a complete
match, we perform \textit{Half Matching} based on the best match of ERPs
or SRPs, respectively. This part of SynAlign, detailed in Listing~\ref{listing:fullHalfMatch},
is called Full+Half Match.

As illustrated in Listing~\ref{listing:SynAlign_algo},
Sequential nets (driver and sink nets of sequential nodes) are aligned first
after Anchor Point Matching, with combinational nets temporarily removed from
$G_{ref}$ and $G_{synth}$. This simplifies the graph to Anchor points and sequential nets
only, which can reduce the problem size to 15\% as in a design
RocketTile~\cite{Asanović:EECS-2016-17}. Reduced sequential graphs retain necessary RP
information for alignment. Before starting the Combinational Full+Half matching
stage, sequential RPs are aligned and both graphs updated, thus reducing the time
complexity.

\begin{center}
  \begin{lstlisting}[style=mystyle, caption={Full+Half Matching algorithm overview}, label={listing:fullHalfMatch},escapechar=|]
for(|$N_s : pending \; nets \; of \; G_{synth}$|) {
  |$SRP\_N_s \;=\; SRPs\; of\; N_s$|
  |$ERP\_N_s\; =\; ERPs \; of\; N_s$|
  for(|$N_r:\; (G_{ref} \; nets\; with\; SRPs\; ==\; SRP\_Ns$|)) {
    |$SRP\_N_r\; =\; SRPs\;  of\;  N_r$|
    |$ERP\_N_r\; =\; ERPs\;  of\;  N_r$|
    if(|$ERP\_N_r\; ==\; ERP\_N_s$|)
      |$record \; as \; Full \; match \; and \; break$|
    else
      |$match\_wt\; =\; calc\_wt(ERP\_N_s,\; ERP\_N_r)$|
  }
  if(|$Full$| |$match$| |$was$| |$captured$|)
    |$N_s\; =\; N_r$| // record as aligned and break
  else {
    |$N_s\; =\; ( N_r\; with\; maximum\; match\_wt)$|
    |$record\; as\; Half\; match\; for\; ERPs\; and\; break$|
  }
  for(|$Nr: \; G_{ref}$| nets with |$ERPs\: ==\: EP\_N_s$|)
    |$match\_wt\: =\: calc\_wt(SRP\_N_s,\; SRP\_N_r)$|
  |$N_s\; =\; (N_r\; with\; maximum\; match\_wt)$|
  |$record\; as\; Half\; match\; for\; SRPs$|
}
\end{lstlisting}
\end{center}

The datatype-specific edges for both graphs are $m_d$, where $m_d << m$.
Thus, when aligning sequential only nets, $m_d$ would represent the number sequential node edges,
and while aligning the combinational logic, $m_d$ would represent the number of
combinational node edges.
If $a$ nets are aligned in previous stages, then $(m_d - a)$ are the
pending nets in synthesized graph. This further lowers the number of entities to
be considered for alignment.
For $G_{ref}$, only nets with the same SRP as on
$G_{synth}$ are processed, termed as $m_d^\prime$. This calculation of
$m_d^\prime$ reduces the \textit{inner for loop} of Listing~\ref{listing:fullHalfMatch}
to 8\% in designs like RocketTile.
Therefore, the total time complexity of the
current matching stage is $O((m_d - a) * m_d^\prime)$.

\subsection{{\bf Partial Matching}} \label{subsec:partialMatch}

If the previous stage does not resolve the remaining nets of the targeted datatype,
\textit{Partial Matching} is performed.
This process is also divided into two
stages, depending on the datatype being processed. 
First, if all sequential
net names are not aligned by line~\ref{line:seqFullHalfMatch} in Listing~\ref{listing:SynAlign_algo},
\textit{Partial Matching} for the remaining sequential net names is carried out. Second,
if line~\ref{line:surrMatch} in Listing~\ref{listing:SynAlign_algo} does not complete the alignment
of annotated net names, \textit{Partial Matching} for annotated combinational
net names is performed.
The goal is to get the best possible match for remaining
unaligned net names. We formulated a $calc\_wt()$ function based on experiments
with benchmarks PipelinedCPU and SingleCycleCPU~\cite{PipelinedCPU,SingleCycleCPU},
where $match\_wt$ is directly proportional to the number of matches.

The overview of the Partial Matching algorithm is in Listing~\ref{listing:partSeq}.
Similar to the time complexity calculation in Section~\ref{subsection:fullHalfMatch},
Partial Matching time complexity computes to $O( (m_d - a) * m{}_d^\prime )$, where
$m{}_d^\prime$ represents edges with common inputs, and $a$ is updated
as per previous matches.
Partial Matching is done only if some annotated node is yet to be resolved. If
only one node is pending resolution and annotated, then $(m_d - a)$ equals one.

\begin{center}
\begin{lstlisting}[style=mystyle, caption={Algorithm overview for Partial matching}, label={listing:partSeq},escapechar=|]
float |$calc\_wt$|(|$synth\_set,\; ref\_set$|) {
  |$num\_matches$| = |$common$| |$entries$| |$in$| |$synth\_set$| |$and$| |$ref\_set$|
  |$mismatches\; =\; synth\_set\; -\; num\_matches $|
  return |$(5*num\_matches)/mismatches$|
}
//Partial Matching:
for(|$N_s : pending \; nets \; of \; G_{synth}$|) {
  for(|$N_r$|: |$G_{ref}\; nets\; with\; atleast\; 1\; SRP\; common\; to\; that\; of\; Ns$|){
    |$SRP\_match\_wt=calc\_wt(SRP\_N_r, SRP\_N_s)$|
    |$ERP\_match\_wt=calc\_wt(ERP\_N_r,ERP\_N_s)$|
    |$match\_wt = SRP\_match\_wt\; + \; ERP\_match\_wt$|
  }
  |$N_s = (N_r$| |$with$| |$maximum$| |$match\_wt)$|
}
\end{lstlisting}
\end{center}

\subsection{{\bf Surrounding Matching}}
\label{subsubsec:surrCell}

\begin{figure}[htpb]
    \centering
        \includegraphics[width=\linewidth]{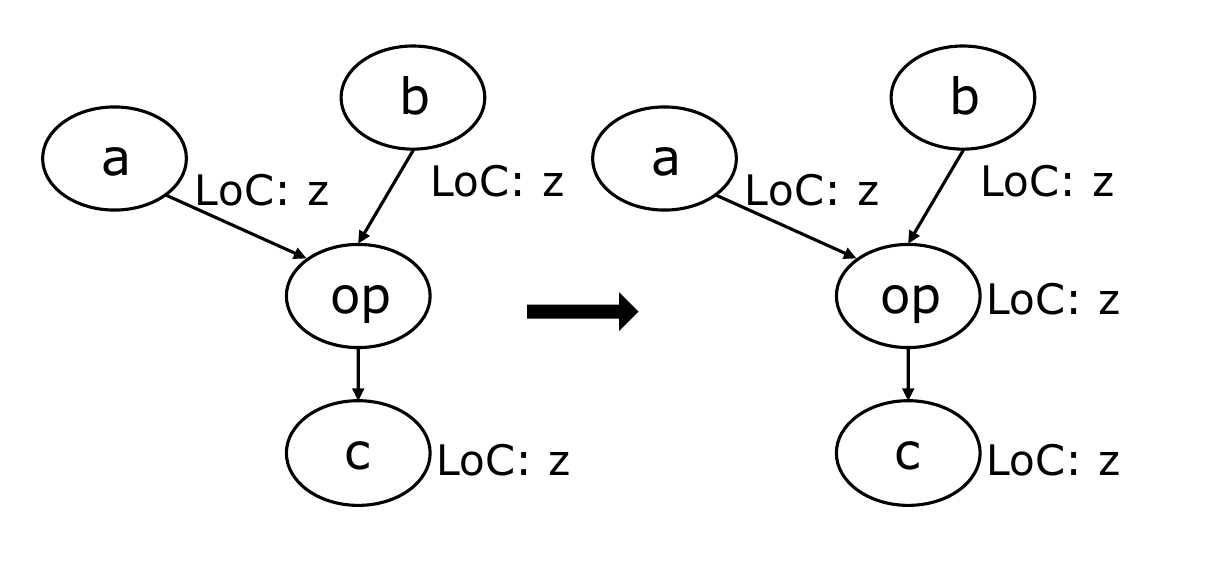}
    \caption{Surrounding Matching depicting the surrounding cells with same LoC
    used to resolve the source location of node ``op''.}
    \label{fig:surr2}
\end{figure}

Since our goal is to reference the source code, we can utilize the Line-of-code
(LoC) information in the $G_{ref}$. Let us define a graph node's directly
connected nodes as Surrounding nodes. If the LoC of all the Surrounding nodes is
the same, then the node and its associated driver RPs can be located
at the same LoC as the surrounding nodes, as illustrated in Figure~\ref{fig:surr2}.
We leverage this information to align more nodes in the $G_{synth}$ with high confidence of alignment.

\begin{figure}[htpb]
  \centering
      \includegraphics[width=.8\linewidth]{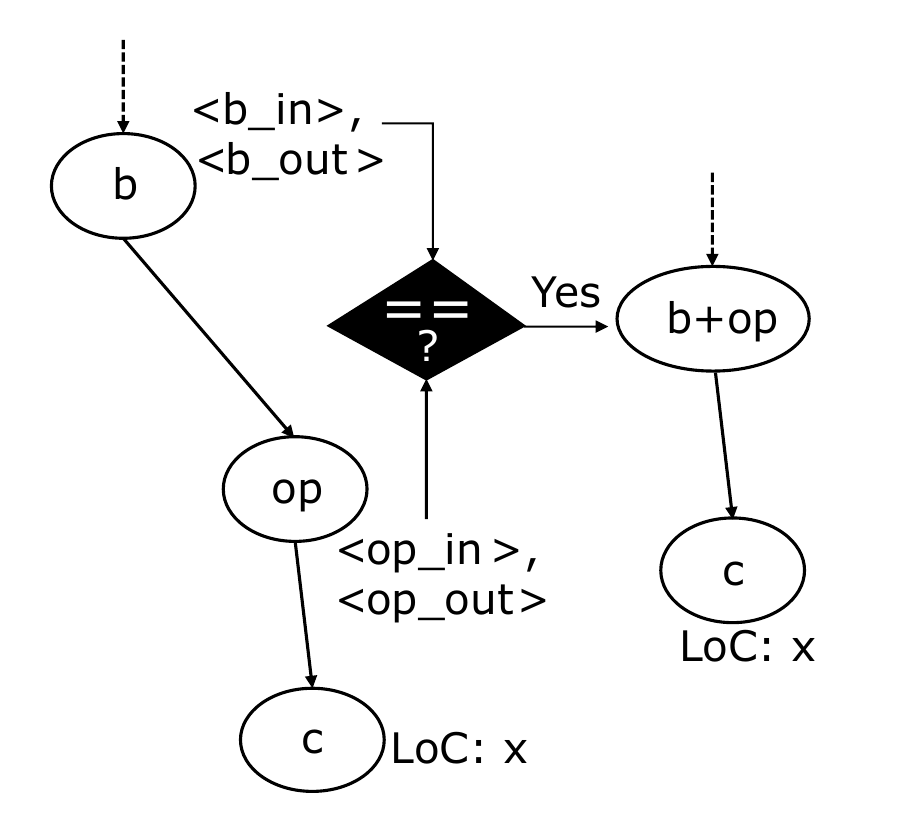}
  \caption{Concept demonstration of Surrounding Matching with collapsing nodes.}
  \label{fig:surr1}
\end{figure}

Another step in Surrounding Matching involves collapsing two nodes, as
illustrated in Figure~\ref{fig:surr1}.
For instance, to find the LoC of the ``op'' node, which is directly connected to ``b''
and ``c'' (with only ``c'' having a known LoC), we check if RPs of ``b'' and ``op''
are the same ($b\_in==op\_in\ \ \&\&\ \ b\_out==op\_out$). If true, nodes
``b'' and ``op'' can be collapsed into a single node and again undergo
\textit{Surrounding Matching}(Figure~\ref{fig:surr2}).

Given its intuitive nature, \textit{Surrounding Matching} is expected to be more accurate
than partial matching stages. However, it is performed later to leverage
more resolved surrounding points. The time complexity of current stage
is $O( 2 * k * (m-a))$, where $k$ (the average number of nodes directly connected
to any node) is a constant, determined to be 5. The factor of 2 accounts for
node collapsing, as shown in Figure~\ref{fig:surr1}.

\begin{table}[h!]
\centering
\caption{Time complexities of different alignment stages.}
\label{timeComplexityTbl}
\begin{tabular}{|l|l|}
\toprule
\textbf{Stage} & \textbf{Complexity}\\
\midrule
Anchor points capturing & $2*O(m)$ \\
Full+Half matching & $O((m_d - a) * m_d^\prime)$ \\
Partial Matching & $O( (m_d - a) * m{}_d^\prime )$ \\
Surrounding Node Matching & $O( 2 * k * (m-a))$ \\
\bottomrule
\end{tabular}

\end{table}

These common steps result in a time complexity less than quadratic, making
SynAlign more efficient than traditional NP network alignment
problems~\cite{zhang2016final}. Unlike SynAlign, general alignment problems
typically have quadratic~\cite{zhang2016final} or cubic~\cite{heimann2018regal}
complexities. Table~\ref{timeComplexityTbl} summarizes the time complexities of
all SynAlign matching stages. The complexity of each step is detailed below.

Thus, SynAlign iteratively aligns the synthesized graph with the reference graph
while keeping the problem size to a minimum in each iteration.

\section{Evaluation and Discussion}\label{sec:eval}

To demonstrate the independence of our research from specific synthesis tools,
we evaluate our work using both an open-source compiler, Yosys~\cite{yosys},
and an industrial compiler, DC~\cite{synopsys:design-compiler}, with Skywater130
and SAED32 liberty technology files, respectively.

SynAlign provides a Line of Code (LoC) from $G_{ref}$ for any annotated $G_{synth}$,
but even careful manual inspection cannot easily confirm its accuracy.
To address this, we propose two evaluation methods: Netlist-to-Netlist Evaluation and Manual Evaluation. 
The benchmarks used in these methods are listed in Table~\ref{tbl:bmSizes}.

\begin{table}[h!]
\caption{Benchmarks' top module names with the corresponding Synthesis compiler and
the generated netlist size (both in k gates and the precise number of gates).}
\label{tbl:bmSizes}
\centering
\begin{tabular}{lcc}
\toprule
\textbf{Benchmark name}&\textbf{Netlist}&\textbf{Netlist}\\
\textbf{(Yosys compiled)}&\textbf{Size (k gates)}&\textbf{Size}\\
\midrule
Mac~\cite{Shashank_06_mac} & 1 & 1210 \\
SingleCycleCPU~\cite{SingleCycleCPU} & 17 & 17305\\
PipelinedCPU~\cite{PipelinedCPU} & 20 & 19596\\
Ibtida~\cite{ibtida} & 35 & 35494\\
Marmot~\cite{marmot_asic} & 84 & 83809\\
RocketTile~\cite{Asanović:EECS-2016-17} & 117 & 116550\\
\bottomrule
\end{tabular}

\begin{tabular}{lcc}
\toprule
\textbf{Benchmark name} & \textbf{Netlist}&\textbf{Netlist}\\
\textbf{(DC compiled)} & \textbf{Size (k gates)}&\textbf{Size}\\
\midrule
SingleCycleCPU~\cite{SingleCycleCPU} & 11 & 10613\\
PipelinedCPU~\cite{PipelinedCPU}& 14 & 13972\\
Marmot~\cite{marmot_asic}& 71 & 70784\\
RocketTile~\cite{Asanović:EECS-2016-17}& 94 & 93549\\
UnoptRocketTile~\cite{Asanović:EECS-2016-17}& 99 & 98857\\
\bottomrule
\end{tabular}

\end{table}

\subsection{Netlist to Netlist (NL2NL) Evaluation}
\label{subsec:nl2nl}

NL2NL testing involves running alignment on a netlist against a name-changed version
of itself to have a scriptable and scalable preliminary testing. This process,
detailed in Listing~\ref{nl2nlSetuplist},
uses a design as $V_{ref}$ and creates an annotated $V_{synth}$ by randomly selecting and
renaming a (noise) percentage of nets by appending \_changed to their names. 
In this context, synth\_net refers to nets from $V_{synth}$, and ref\_net represents nets in $V_{ref}$.

To create $V_{synth}$ from $V_{ref}$, we implement a compiler pass to collect 
all net names in the graph and randomly select a percentage (\textit{noise\%}) of these names. 
These selected nets, along with the module names, are then suffixed
with ``\_changed'', thus creating the $V_{synth}$ for NL2NL Evaluation. Subsequently, 
$V_{synth}$ and $V_{ref}$ (one with ``\_changed'' and one without) 
are used for alignment.

\begin{figure}[htpb]
    \centering
        \includegraphics[width=1\linewidth]{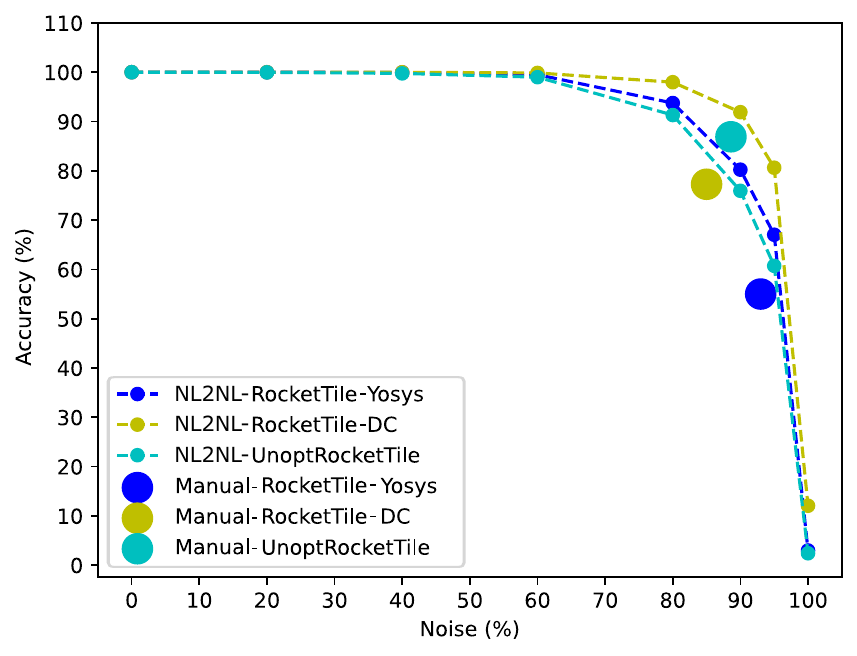}
    \caption{NL2NL accuracy Plot for different RocketTile compilations}
    \label{fig:nl2nlAccPlot}
\end{figure}

\begin{figure}[htpb]
    \centering
        \includegraphics[width=1\linewidth]{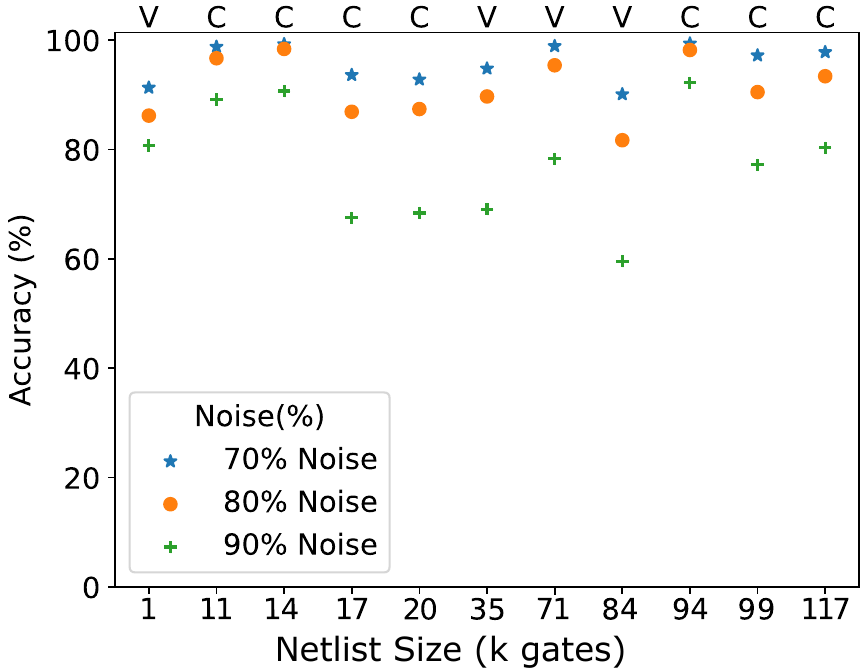}
    \caption{NL2NL accuracy testing in all the Verilog (V) and Chisel (C) designs}
    \label{fig:nl2nlVerPlot}
\end{figure}

The alignment results are used to assess accuracy, as
depicted in Figure~\ref{fig:nl2nlAccPlot}, for different RocketTile synthesis
options from Table~\ref{tbl:bmSizes}. At 0\% noise, all net names are preserved,
resulting in 100\% accuracy due to Anchor Points matching. As noise increases, SynAlign maintains 100\%
matching accuracy until about 60\% noise.
The three lines corresponding to RocketTile show similar trends.
For brevity, we do not display individual curves for each benchmark, but all
follow similar patterns, as reflected in Figure~\ref{fig:nl2nlVerPlot}.
Figure~\ref{fig:nl2nlVerPlot} shows 
data points at 70\%, 80\%, and 90\% noise levels for all the designs in Table~\ref{tbl:bmSizes}, indicating
consistent performance across various designs, languages, and synthesis tools.
This figure also indicates that SynAlign can tolerate
significant noise levels in netlists.
Figure~\ref{fig:nl2nlVerPlot} header denotes whether the source
code is Verilog (\textit{V}) or Chisel (\textit{C}).

{\linespread{1.15}\selectfont
\begin{lstlisting}[style=mystyle, numbers=none, caption={NL2NL experimental setup overview}, label={nl2nlSetuplist},escapechar=|]
for (|$noise\;:\; [0,20,40,60,80,90,95,100]\% \;of\; V_{synth}\; nets$|) {
  |$V_{synth}\;=\;V_{ref}\; with \;  noise\% \;  nets \;  appended \;  with \;  ``\_changed"$|
  |$Perform   \; alignment.$|
  if (|$synth\_net.erase(``\_changed")\; == \; ref\_net$|) 
    |$record\; as\; matched.$|
  else 
    |$record\; as\; mismatched.$|
  |$\smash{accuracy = \frac{{number  \, of  \, entries  \, matched}}{{total  \, number  \, of  \, entries}}}$|
}
\end{lstlisting}
}

Moreover, when sequential net names are preserved, and all other
combinational net names are randomized, designs like SingleCycleCPU
and PipelinedCPU can tolerate up to 100\% noise with 100\% accuracy.
This suggests that SynAlign is more sensitive to noise at the sequential
points than at the combinational points.

NL2NL testing provides valuable insights into the algorithm's
performance in a controlled environment, facilitating automated scalability
testing. Thus, this approach is more manageable compared to the Manual Evaluation
discussed next.

\subsection{Manual Accuracy Evaluation}
\label{subsec:manualEval}

While NL2NL evaluation offers valuable insights, it does not account for
structural changes introduced by synthesis tools.
Hence, we set up the Manual Accuracy Evaluation. It is performed
for larger benchmarks from Table~\ref{tbl:bmSizes}.
To validate that a synthesized node aligns with its source code,
we manually verify if a node in the netlist corresponds to the intended LoC.
Knowing the correct match beforehand is crucial, so we preserve only the variable
to be aligned for this evaluation.
In synthesized designs,
many graph nodes and edges may be eliminated or duplicated due to synthesis. To address this, we
manually mark certain variables in the source code as ``dont\_touch'' or
``keep'' before synthesis.

\begin{figure}[htpb]
\centering
    \includegraphics[width=\linewidth]{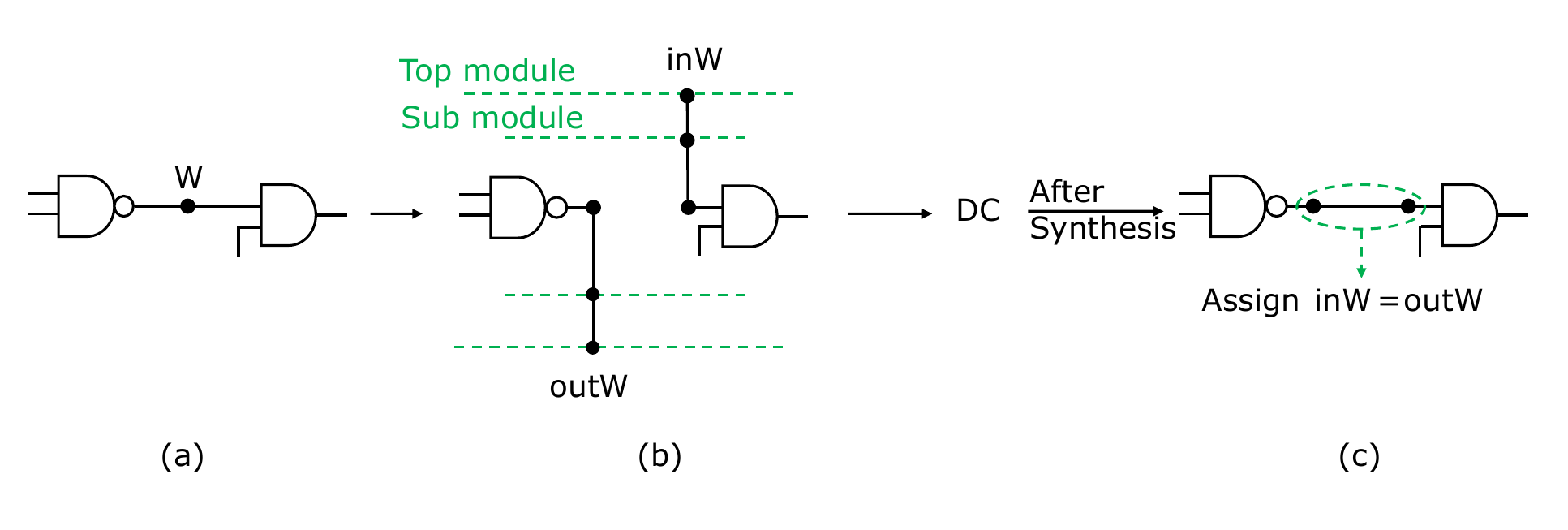}
\caption{(a) Original net to be preserved (W), (b) punched net, (c) Punched nets connected back}
\label{punchthrough}
\end{figure}

Manual evaluation process involves the following steps:
\begin{enumerate}
    \item {\textbf{Script for random marking:} Write a script to randomly select
    multiple Chisel-assignment
    lines in the source code, ensuring these lines are within the Chisel-emitted
    Verilog. The chisel assignment
    lines with constant right-hand side (RHS) should not be selected.
    Mark the Left-Hand Side (LHS) of the selected lines as
    `dontTouch'~\cite{lund2021design} for manual validation.}
    \item {\textbf{Emit Reference Verilog ($V_{ref}$):} HDL compilers support Verilog generation with
    source LoC information~\cite{getMeVerilogChisel,RocketChipGenerator}. This
    generated Verilog, with the `dontTouch'
    annotations, serves as $V_{ref}$ in Chisel benchmarks. In Verilog
    benchmarks, the source code itself serves as $V_{ref}$. }
    \item {\textbf{Synthesized Verilog ($V_{synth}$):} Use $V_{ref}$ to create $V_{synth}$.
    Ensure synthesis does not remove `dontTouch' variables by marking them with
    `keep'~\cite{wolf2021yosys,yosysKeep} in Yosys.
    
    For DC, use `set\_dont\_touch' and implement punching through the design
    (as illustrated in figure~\ref{punchthrough}) to preserve variables/nets through the synthesis cycle.
    Punching is required in DC to ensure  
    the `set\_dont\_touch' directive is not overwritten during compilation. 
    This is necessary because $V_{ref}$, being an HDL verilog, contains generic logic.

    To achieve maximum optimization of the design, enable flattening and retiming in all the synthesis runs.
    To test the impact on unoptimized synthesized designs, we use the benchmark
    UnoptRocketTile. Its synthesized netlist was obtained from DC without enabling
    retiming and flattening, making the UnoptRocketTile a hierarchical, unoptimized benchmark.
    }
    \item {\textbf{Annotation and Alignment:} Annotate $V_{synth}$ using a JSON
    file with the module and node names.
    Additionally, the annotated net name is manually changed in the netlist to
    prevent Anchor Point matching of the same.
    The alignment algorithm detects the annotation attribute on nodes and marks them
    for matching against the source code. The result is printing of the annotated node name,
    its aligned $G_{ref}$-node name, and the LoC information from $V_{ref}$.
    
    This annotation and alignment process is carried out
    using an open-source framework called LiveHD~\cite{livehd}.
    By reading this information, an engineer can easily identify the location of the
    annotated cell in the source code.
    }
    \item {\textbf{Aligned source LoC:} If the source HDL is not Verilog, a LoC
    containing $V_{ref}$ is emitted for
    alignment. The LoC information printed during the alignment step can be used to trace back to the
    original HDL source code.}
\end{enumerate}

\begin{figure}[htpb]
\centering
    \includegraphics[width=1\linewidth]{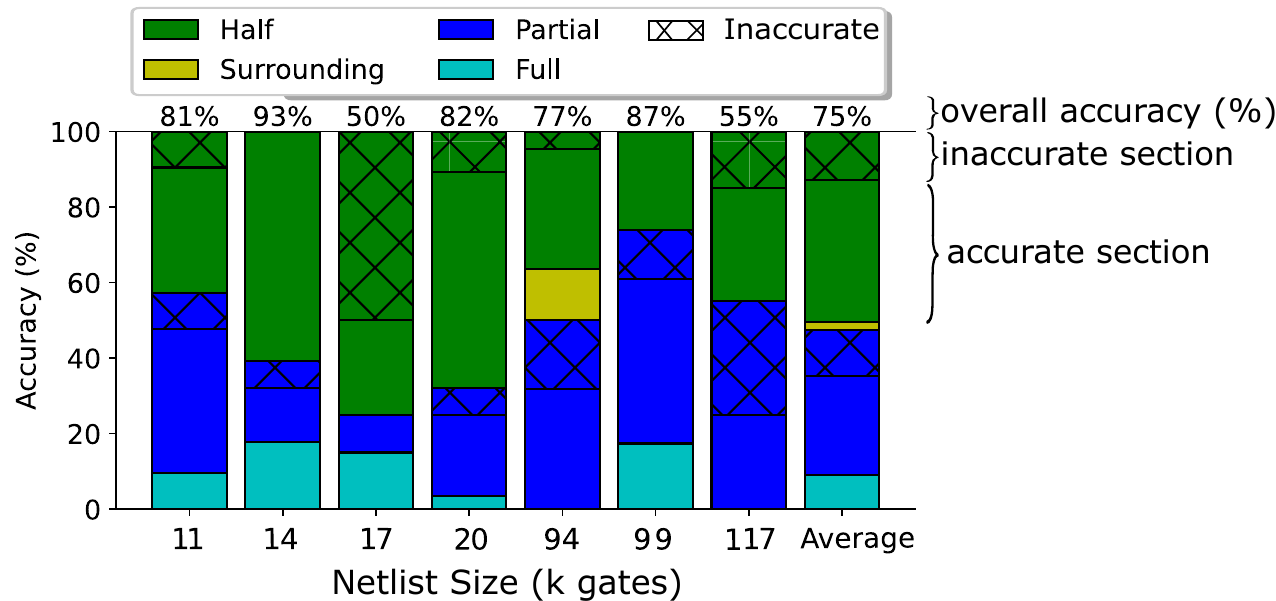}
\caption{Plot depicting the breakdown of different matching functions with the total benchmark accuracy on top of the bars.
Sequential and combinational partial matching is combined as Partial.
Similar case for Full+Half matching functions.}
\label{fig:manualAccPlot}
\end{figure}

\begin{figure*}[htpb!]
\centering
    \includegraphics[width=1\linewidth]{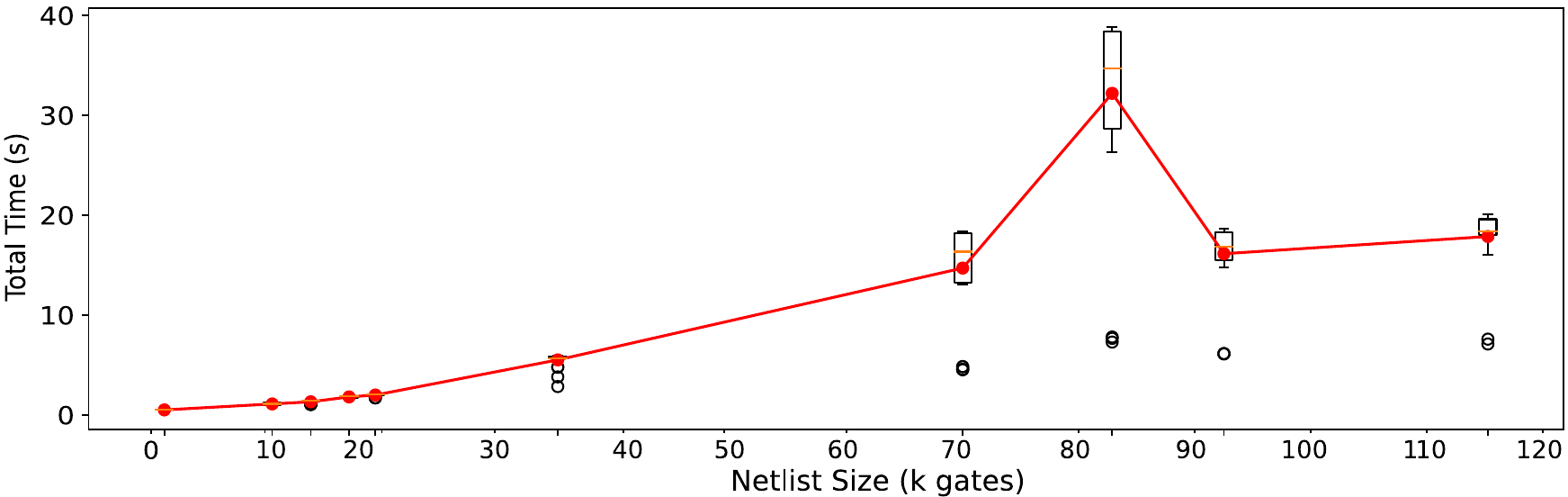}
\caption{Total Execution Time Variation as per Netlist Size.}
\label{TotalExecutionTimeVariationAsPerNetlistSize}
\end{figure*}

Figure~\ref{fig:nl2nlAccPlot} shows manual marking accuracies, with the x-axis derived
from the percentage of aligned Anchor points.
For Manual-RocketTile-Yosys, this corresponds to 90-95\% on the NL2NL
noise axis, with a similar calculation for RocketTile\_DC (Manual-DC).
Manual evaluation accuracy is lower than NL2NL due to synthesis-induced structural changes
and LoC precision loss in Verilog “always blocks” (Section~\ref{section:insight}).
Manual-UnoptRocketTile represents a DC-generated netlist without flattening and
retiming. It shows higher accuracy due to lower optimization levels, which
preserve more of the original structure. It also preserves more sequential flop
names: 85\% compared to 79\% in Manual-DC. SynAlign performs significantly better
with higher preservation of sequential nodes,
indicating consistency between NL2NL and Manual evaluation.

Figure~\ref{fig:manualAccPlot} breaks down the accuracy
for each stage from section~\ref{sec:methodology}. 
Crossed part of the bar (marked ``X'') in this figure
indicates the percentage of inaccurately matched annotated nets.
As an example, in
figure~\ref{fig:manualAccPlot}, RocketTile\_DC (from Table~\ref{tbl:bmSizes})
performs alignment of the annotated nodes in \textit{Half},
\textit{Surrounding}, and \textit{Partial} matching functions. Around 50\% of
matching is done during \textit{Partial Matching}, out of which the accurate
matches are 30\%, and ``X'' represents the rest with the incorrect alignment.
We can observe that Full and Surrounding matches have 100\% accuracy across all benchmarks,
while Half and Partial Matching have an average error of around 25\%.
Thus, alignment during Anchor Point, Full, and Surrounding Matching can be highlighted in
SynAlign result as most confident match.
Manual accuracy ranges from 50\% to 93\% across languages, designs, and synthesis
optimizations. Also, lower optimization
levels meaning more Anchor points, show higher accuracy.

These combined evaluation methods demonstrate the
robustness and accuracy of SynAlign across different designs and synthesis
tools, highlighting its capability to tolerate significant noise
levels in netlists.

\begin{figure}[htpb]
\centering
    \includegraphics[width=1\linewidth]{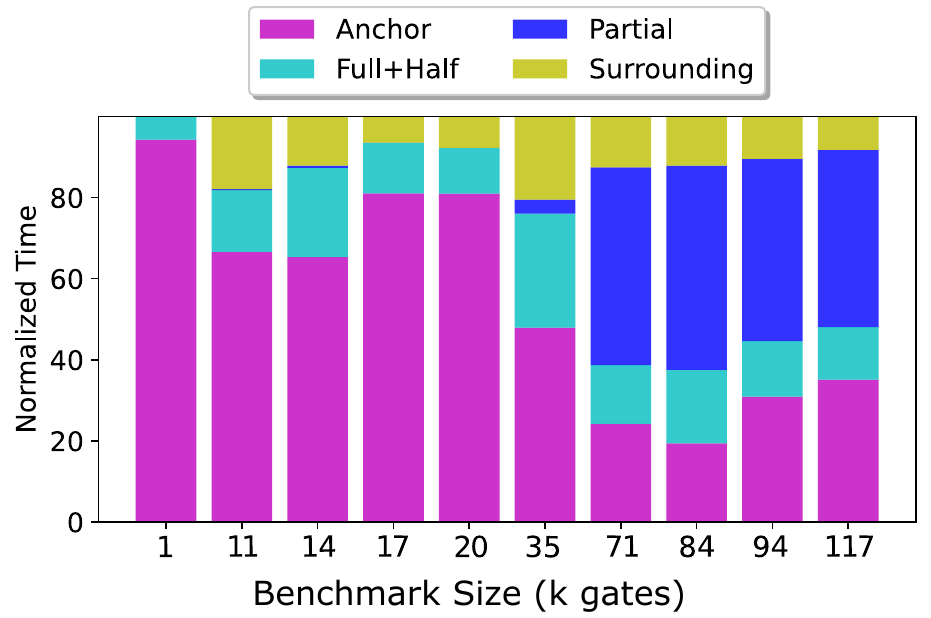}
\caption{Breakdown of different matching functions for all the Benchmarks.}
\label{TimeBreakdownPlot}
\end{figure}

\begin{figure}[htpb!]
\centering
    \includegraphics[width=1\linewidth]{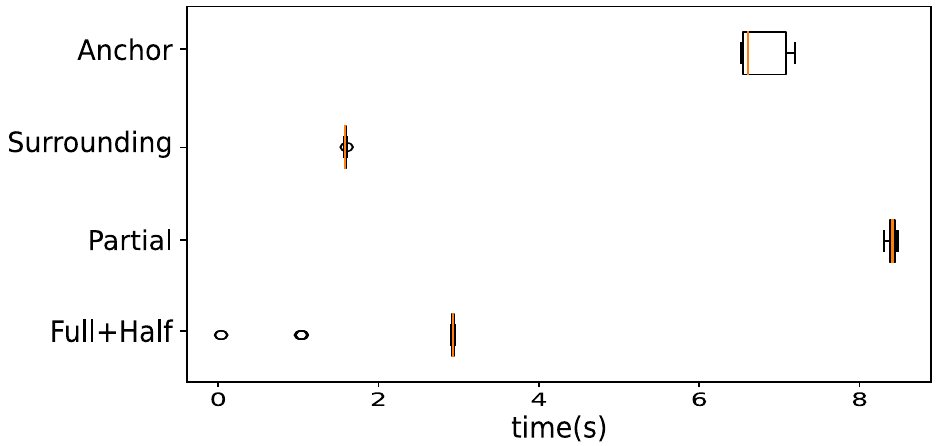}
\caption{Average time taken by different matching functions for random net alignment in RocketTile\_Yosys.}
\label{TimeVsBreakdownForRocketTile}
\end{figure}

\begin{figure}[htpb]
    \centering
        \includegraphics[width=1\linewidth]{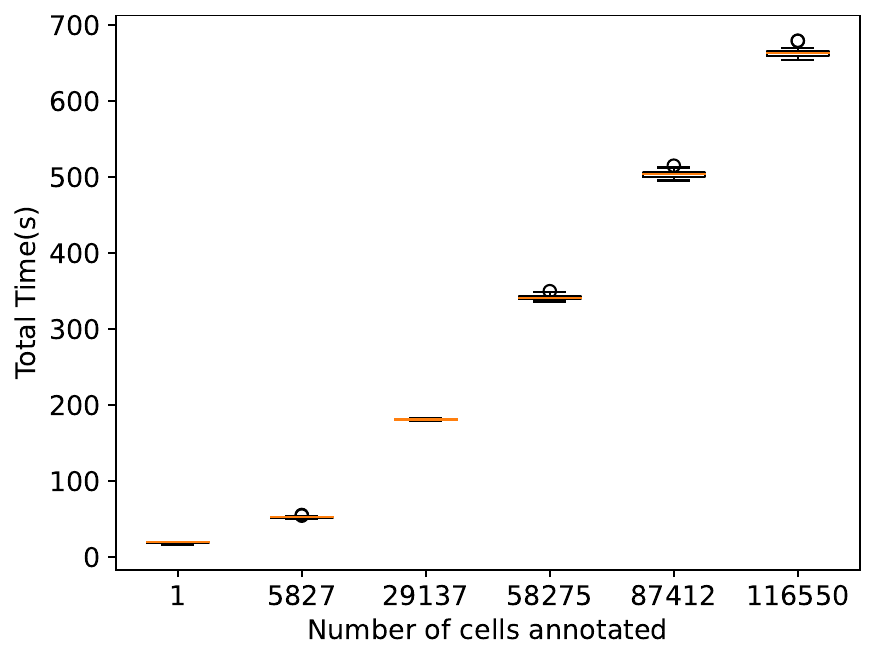}
    \caption{Time variation as per number of net names annotated for the Design RocketTile synthesized using Yosys.}
    \label{TimeVariationAsPerNumberOfCellsAnnotatedForRocketTileYosys}
\end{figure}

\subsection{Performace evaluation}
\label{subsec:perfEval}

To evaluate SynAlign's performance, each benchmark is run 30 times, and a
different node is annotated each time. Unlike previous evaluations, net
names need not be changed. 

SynAlign's runtime
is divided into four main stages mentioned in Section~\ref{sec:methodology}.
Figure~\ref{TimeBreakdownPlot} shows that as netlist size
increases, Partial Matching consumes most of the time, while Full+Half
Matching consistently takes 10-20\% of the time, regardless of netlist size
and compiler used. This is due to the algorithmic optimization
discussed in Section~\ref{subsection:fullHalfMatch}. Figure~\ref{TimeVsBreakdownForRocketTile} presents
the average runtime in seconds for random net name annotations in each benchmark.
It highlights that a trivial function variation
can occur depending on the node annotated.
This is reflected in Full+Half Matching time variation in the figure.

Scalability is illustrated in Figure~\ref{TotalExecutionTimeVariationAsPerNetlistSize}.
It shows that total execution time increases linearly with netlist size,
with the Marmot-yosys benchmark taking significantly longer due to its
$V_{ref}$ being six times larger than that of RocketTile.
The results demonstrate that SynAlign's performance scales with problem size.
These three figures focus on single net name
annotations in each run.

The impact of annotating more than one cell was tested on the
largest design in Table~\ref{tbl:bmSizes}, i.e. RocketTile compiled with
Yosys, with random selection varying from one cell to 100\% of cells
of $V_{synth}$. 
It was found to be almost linearly increasing.
The results can be seen in
Figure~\ref{TimeVariationAsPerNumberOfCellsAnnotatedForRocketTileYosys}.
Notably, the total time for single net name annotation in
Figure~\ref{TimeVariationAsPerNumberOfCellsAnnotatedForRocketTileYosys} matches
the total time consumed by the design in
Figure~\ref{TotalExecutionTimeVariationAsPerNetlistSize}.
The results also show that the
performance scales with problem size.

\begin{figure*}
    \begin{lstlisting}[style=mystyle, numbers=none, caption={Excerpt from the log file in Section~\ref{ex2wrong}}, label={ex2log}]
             Annotated node       :      net name in Vref                      : LoC in Vref
dcache.tlb_pmp_io_addr_31_CHANGED : <unnamed edge>(module DCache)              : [26]
dcache.tlb_pmp_io_addr_31_CHANGED : tlb__deny_access_to_debug_T(module DCache) : [70]
dcache.tlb_pmp_io_addr_31_CHANGED : <unnamed edge>(module DCache)              : [70]
\end{lstlisting}
\end{figure*}

\begin{figure*}
\begin{lstlisting}[style=mystyle, numbers=none, language=Verilog, caption={
    Relevant $V_{ref}$ lines with the line numbers for the example in section~\ref{ex2wrong}
    }, label={ex2Vref}]
26 wire [2:0] tlb_mpu_priv = tlb__mpu_priv_T ? 3'h1 : tlb__mpu_priv_T_2; //@[TLB 233]
70 wire tlb__deny_access_to_debug_T = tlb_mpu_priv <= 3'h3; //@[TLB 245]
\end{lstlisting}
\end{figure*}

\begin{figure*}
    \begin{lstlisting}[style=mystyle, numbers=none, caption={Excerpt from the result in~\ref{ex1wrong}.
        The annotated node is aligned with the line number 5047 in Listing~\ref{potatoVref}
    }, label={potatolog}]
     Annotated node         :  net name in Vref              :  LoC in vref
fpuOpt.fpiu.potato2_CHANGED : <unnamed edge>(module FPToInt) : [5047]
\end{lstlisting}
\end{figure*}

\begin{figure*}
    \begin{lstlisting}[style=mystyle, numbers=none, caption={Excerpt from the log file for the
        walkthrough of alignment analyzed as accurate, in section~\ref{ex1right}
    }, label={ex3log}]
        Annotated node                    :     net name in Vref         :  LoC in Vref
frontend.tlb.io_ptw_req_bits_valid_CHANGED : tlb_io_kill(module Frontend) : [132882]
\end{lstlisting}
\end{figure*}

In Section~\ref{sec:methodology}, we highlighted that SynAlign's
overall time complexity is less than quadratic, which is an improvement over
the general network alignment algorithms like~\cite{kazemi2016network,
bayati2013message,zemlyachenko1985graph,kobler2012graph}.
Overall, SynAlign provides results in seconds for large designs like RocketTile
with few net name annotations, a common scenario for designers tracing
connections from a small set of cells to the original source code.

\subsection{Insights to evaluation} \label{section:insight}

In this section, we discuss a few examples of how we evaluated any alignment result as inaccurate or accurate
in Figure~\ref{fig:manualAccPlot}.

\subsubsection{The first walkthrough of alignment analyzed as inaccurate.} \label{ex2wrong}

In the RocketTile (Yosys compiled) benchmark, one of the bits of ``tlb\_pmp\_io\_addr''
was annotated with the help of a JSON file, as mentioned in Section~\ref{subsec:manualEval}.
The precise and expected output in this case would align with line numbers 232, 235, and 236
in the source code shown in Listing~\ref{ex2src}.

\begin{lstlisting}[style=mystyle, numbers=none, language=Verilog,
caption={Source code lines for the example in Section~\ref{ex2wrong}}, label={ex2src}]
232 val mpu_physaddr = Cat(mpu_ppn, io.r...
233 val mpu_priv = Mux[UInt](Bool(usingVM)&&...
234 val pmp = Module(new PMPChecker(lgMaxSize))
235 pmp.io.addr := mpu_physaddr
236 dontTouch(pmp.io.addr)
245 val deny_access_to_debug = mpu_priv <= ...
\end{lstlisting}

The alignment result in Listing~\ref{ex2log} shows the annotated net
along with its aligned nets and source code location. 
From this result, we can see that the node annotated was renamed using ``\_CHANGED''
as explained in the evaluation strategy earlier. Further, the annotated node is aligned
to line numbers 26 and 70 in $V_{ref}$. Thus, to obtain original source code, the
output can be traced to $V_{ref}$ shown
in Listing~\ref{ex2Vref} and further to the source code in Listing~\ref{ex2src}. 
Therefore, for non-Verilog HDLs, source location information in the reference
Verilog is used to trace back to the original source code. Since
SynAlign points to LoC other than the expected output locations, we capture this alignment result
as inaccurate. Even so, we can see that the correct location is nearby and logically
related to the annotated position. 
It is evident that the matching occurred in the
line adjacent to the accurate source.
Therefore, there is significant potential for improving the accuracy of SynAlign.

\begin{lstlisting}[style=mystyle, numbers=none, mathescape=true, caption={
    Excerpt from source code preserved using dontTouch. We made a new variable called \textit{potato2}
    for the example in~\ref{ex1wrong}}, label={potatoSrc}]
when(cvtType==i){...}
      $\downarrow$
val potato2 = (cvtType==i)
when(potato2){...}
dontTouch(potato2)
\end{lstlisting}

\subsubsection{The second walkthrough of alignment analyzed as inaccurate.} \label{ex1wrong}

As in the first walkthrough (Section~\ref{ex2wrong}), we
annotate another node in the RocketTile (Yosys compiled) benchmark
and evaluate the alignment in this second walkthrough.
As explained in section~\ref{subsec:manualEval}, we applied \textit{dontTouch}
on a while loop condition. The corresponding source code is presented in Listing~\ref{potatoSrc}.
In order to evaluate the alignment, the condition ``cvtType==i'' was preserved
in a random variable named ``potato2'', as shown in Listing~\ref{potatoSrc}.

The alignment result for this preserved net is shown in
Listing~\ref{potatolog}. This listing indicates that the annotated node points to line number 5047 in
$V_{ref}$. $V_{ref}$-line 5047, as seen in Listing~\ref{potatoVref}, points to the beginning of an ``always'' block. 
This ``always'' block connects to the annotated variable via a register ``in\_typ''. 
However, ``in\_typ'' is not mentioned under the aligned \textit{net name in $V_{ref}$} in Listing~\ref{potatolog},
leading us to consider this matching as
inaccurate. For the current annotation to be recognized as an accurate match, the resultant Listing~\ref{potatolog}
should have mentioned either of 4920, 4919, or 5061 $V_{ref}$ line numbers.

\begin{lstlisting}[style=mystyle, numbers=none, language=Verilog, caption={Relevant lines with the line numbers from $G_{ref}$ Verilog output
    generated using Listing~\ref{potatoSrc}. The LoC information is also available in $G_{ref}$ as shown here.
    }, label={potatoVref}]
4919 wire cvtType = in_typ[1]; //@[package 154]
4920 wire potato2 = ~cvtType; //@[FPU 501]
5047 always @(posedge clock) begin
     ...
5061    in_typ <= io_in_bits_typ; //@[Reg 17]
     ...
5072 end // end of always block
\end{lstlisting}

\begin{lstlisting}[style=mystyle, numbers=none, language=Verilog, caption={Source code lines for the
    walkthrough of alignment analyzed as accurate, as mentioned in section~\ref{ex1right}
    }, label={ex3src}]
447   io.ptw.req.bits.valid := !io.kill
452   dontTouch(io.ptw.req.bits.valid)
\end{lstlisting}

\subsubsection{A walkthrough of alignment analyzed as accurate.} \label{ex1right}

Now we discuss an example of a positive accuracy analysis used for Figure~\ref{fig:manualAccPlot}.
The annotated node was ``frontend.tlb.io\_ptw\_req\_bits\_valid\_CHANGED'', and SynAlign aligned this
node with ``tlb\_io\_kill'' from the module Frontend, as can be seen in Listing~\ref{ex3log}.
Referring to the source code in Listing~\ref{ex3src}, it is evident
that the annotated variable and the $G_{ref}$ aligned variable are connected by single 1-input logic gate.
Therefore, this alignment is considered a success and is marked as an accurate match.

\section{Limitations and Future Work}
\label{sec:future}

This work relies on the availability of Anchor links. The presence of more
Anchor links could significantly enhance reliability, and conversely, their
absence could reduce it.

In sections~\ref{ex1wrong} and \ref{ex2wrong},
we observed that alignments marked as inaccurate were often
indirectly connected to or near the preserved variable.

Open-source tools like LiveHD with Yosys frontend do not capture precise LoC. For instance,
in Listing~\ref{listingAlwaysBlock}, an ``always''
block in the Verilog source code on line number 241 translates to 122 registers
in the netlist notation via Yosys, all pointing to the beginning of the
``always block'', as shown in Listing~\ref{listingAlwaysLG}.
This results in a loss of data when pinpointing the exact source location.
Preserving the LoC for different flip-flops in
Listing~\ref{listingAlwaysLG} would improve SynAlign's accuracy.

\begin{lstlisting}[style=mystyle, language=Verilog, numbers=none, 
    caption={Always block in RocketTile's Source Verilog}, label={listingAlwaysBlock}]
241 always @(posedge clock) begin
...
248 id_reg_pause <= _GEN_1;
...
252 ex_ctrl_fp<=id_ctrl_dec;//@[RocketCore 445
\end{lstlisting}

\begin{lstlisting}[style=mystyle, numbers=none,caption={Some netlist nodes for Listing~\ref{listingAlwaysBlock}}, label={listingAlwaysLG}]
nid:715 type:flop module:Rocket loc:[241,0]
nid:716 type:flop module:Rocket loc:[241,0]
nid:717 type:flop module:Rocket loc:[241,0]
\end{lstlisting}

Furthermore, we envision a GUI-based tool to facilitate chip design engineers in
leveraging this work, as illustrated in
Figure~\ref{fig:annotate}. For example, integrating SynAlign with tools like 
Verdi~\cite{verdi,maidhili2020reset,ghosh2019case} would be beneficial. While
this is an intriguing direction that could enhance the impact of our work, it is
beyond the scope of this study.

\section{Related Work}\label{sec:related}

Tools like Formality verify functional equivalence by
analyzing how inputs propagate through the design
and ensuring the outputs match between RTL and gate-level
netlists. While effective for functional verification,
this method lacks a clear, user-friendly mapping of
specific RTL elements to their netlist counterparts.
This limitation reduces transparency in the
debugging process and requires additional manual
effort to trace synthesis transformations.

In contrast, SynAlign provides a detailed point-to-point
alignment between the RTL and netlist, mapping each element
directly to its post-synthesis equivalent. This detailed
mapping offers greater insight into how synthesis
optimizations affect specific RTL components,
reducing manual intervention and speeding up
debugging and analysis. This makes SynAlign superior
for understanding the design flow compared to traditional
equivalence tools that rely primarily on logic cone
analysis~\cite{lec_formality_design_reuse}.

\subsection{HDL Compilers}\label{subsec:hwCompiler}

Many HDL compilers, such as Chisel~\cite{chisel}, XLS~\cite{xls}, and CIRCT~\cite{circt},
emphasize the importance of propagating source code locations through their compiler
passes. XLS highlights that maintaining source correlation enhances debugging,
visualization, and productivity~\cite{IdeasXLS}. However,
these compilers often rely on external synthesis tools that discard the
source locations preserved by the HDL compilers.

CIRCT developers also stress the importance of location
tracking~\cite{CIRCT-Notes, CIRCT-Notes1}, using LLVM infrastructure to
maintain source locations. Accessing these details from LLVM IR involves
navigating a complex hierarchy~\cite{srcLoc_2021}. Each transformation
in these compilers must maintain instruction debug locations, which is both
complex and resource-intensive~\cite{LLVMdocumentation, Update_Debug_Info_LLVM}.
Modifying or developing passes requires numerous considerations for source
location~\cite{Update_Debug_Info_LLVM1}.

In contrast, SynAlign does not impose restrictions on
developers or require efforts to maintain source-level debugging information.
No modifications to compiler passes or transformations are needed for debug
information with SynAlign.

Recent work on equality saturation~\cite{smith2024there, pal2023equality} explores
its application to EDA tasks but does not provide a line-by-line mapping between the
netlist and the original source code.

This paper is the first to demonstrate how to connect a
post-synthesis netlist with the original source code without propagating
source locators.

\subsection{Network Alignment}\label{subsec:nwAlignment}

Network alignment, which involves matching two graph representations,
is used in various fields such as protein-protein interaction analysis
and cross-platform social network recommendations~\cite{kazemi2016network}.
Traditional approaches often perform one-to-one node matching using bipartite
graphs~\cite{bayati2013message} and evaluate isomorphic networks with weighted
edges~\cite{zemlyachenko1985graph,kobler2012graph}. In contrast,
SynAlign utilizes many-to-many matching to compare Verilog source
code and synthesized netlists based on structural connections and 
explores hardware design attributes.

SynAlign is unique in applying graph alignment to hardware compilers,
addressing the significant complexity and noise introduced by synthesis.
Previous works like~\cite{bayati2013message, zhu2022caper, heimann2021refining}
consider 5\% noise significant~\cite{heimann2021refining}, whereas synthesis
typically introduces over 80\% noise. Unlike~\cite{zhu2022caper},
which uses edge-weighted graphs for group coarsening,
SynAlign's approach is directly applicable in chip design
because synthesis alters the network structure by dropping edges.

Similar to~\cite{liu2016aligning}, SynAlign uses
structural matching with Anchor Links (ports and preserved pin names).
RefiNa~\cite{heimann2021refining} performs structural matching based
on matched neighbourhood consistency (MNC), showing the relationship
between MNC and alignment accuracy. RefiNa operates on graph isomorphism,
treating edge/connection changes as noise. In contrast, SynAlign uses
structural matching based on node datatype or functionality, making it
suitable for highly noisy post-physical implementation design graphs
that would have low accuracy under RefiNa.

SynAlign leverages the advantages of Multinetwork alignment,
as seen in CAPER~\cite{zhu2022caper}, to provide a multi-level solution.
At the coarsest level, it evaluates Anchor links and sequential logic,
followed by combinational logic matching. Section~\ref{sec:methodology}
detailed SynAlign's multi-level approach.

While SynAlign addresses a network alignment problem, its application
and noise properties differ significantly from traditional
network alignment problems. We believe the principles introduced in
this paper can be applied to other areas, such as compiler optimization.
To our knowledge, network alignment has never been used in the chip design
domain to reduce engineering efforts before.

SynAlign represents a novel application of network alignment in the domain
of hardware compilers, specifically addressing the challenges posed by
synthesis-induced noise. By leveraging many-to-many matching and a multi-level
approach inspired by CAPER, SynAlign effectively compares logical and synthesized
netlists based on structural connections and hardware attributes. This innovative
method significantly reduces engineering efforts in chip design, demonstrating the
potential for network alignment principles to be applied in new and complex areas.

\section{Conclusion}\label{sec:conclusion}

Obtaining the source code location for any part of the netlist offers multiple advantages.
Front-end designers can receive early feedback on their designs' timing and power,
allowing them to optimize the source code in advance. Consequently, backend tools can
be provided with mature constraints, reducing the time required for iterative cycles in
chip design and significantly decreasing time-to-market. 

The ability to trace synthesized netlist lines back to the source code,
regardless of the optimization tools used, is crucial for reducing time-to-market
in the chip design industry. SynAlign enables this back-tracing for any HDL
translatable to Verilog with line-of-code (LoC) information without relying on a
single EDA tool. SynAlign achieves an average of 75\% accuracy in less than 20 seconds.

\bibliographystyle{plain}
\bibliography{biblio}

\begin{thebibliography}{10}

\bibitem{getMeVerilogChisel}
{Chisel Users Community, Get Me Verilog}.
\newblock \url{https://www.chisel-lang.org/docs/resources/faqs#get-me-verilog}.
\newblock {Online; accessed on 16 April 2024}.

\bibitem{CIRCT-Notes}
Circt weekly discussion notes.
\newblock
  \url{https://docs.google.com/document/d/1fOSRdyZR2w75D87yU2Ma9h2-_lEPL4NxvhJGJd-s5pk/edit#heading=h.56zyi32ygm3b}.

\bibitem{CIRCT-Notes1}
Circt weekly discussion notes.
\newblock
  \url{https://docs.google.com/document/d/1fOSRdyZR2w75D87yU2Ma9h2-_lEPL4NxvhJGJd-s5pk/edit#heading=h.awos2k698n33}.

\bibitem{clang-locations}
clang: Physical source locations.
\newblock
  \url{https://clang.llvm.org/doxygen/group__CINDEX__LOCATIONS.html#details}.

\bibitem{RocketChipGenerator}
{Generating verilog in Rocket Chip Generator}.
\newblock
  \url{https://github.com/chipsalliance/rocket-chip?tab=readme-ov-file#building-the-project}.
\newblock {Online; accessed on 16 April 2024}.

\bibitem{Update_Debug_Info_LLVM}
How to update debug info: A guide for llvm pass authors — llvm 17.0.0git
  documentation.
\newblock
  \url{https://llvm.org/docs/HowToUpdateDebugInfo.html#when-to-drop-an-instruction-location}.

\bibitem{Update_Debug_Info_LLVM1}
How to update debug info: A guide for llvm pass authors — llvm 17.0.0git
  documentation.
\newblock \url{https://llvm.org/docs/HowToUpdateDebugInfo.html}.

\bibitem{IdeasXLS}
Ideas and projects - xls: Accelerated hw synthesis.
\newblock \url{https://google.github.io/xls/ideas_and_projects}.

\bibitem{netlistsvg}
netlistsvg.
\newblock \url{https://github.com/nturley/netlistsvg}.
\newblock Online; accessed on 21 May 2019.

\bibitem{yosysKeep}
{openLane}.
\newblock
  \url{https://web.open-source-silicon.dev/t/442348/has-anyone-tried-using-the-keep-synthesis-attribute-is-follo}.
\newblock {Online; accessed on 16 April 2024}.

\bibitem{LLVMdocumentation}
Source level debugging with llvm — llvm 17.0.0git documentation.
\newblock \url{https://llvm.org/docs/SourceLevelDebugging.html}.

\bibitem{verdi}
verdi.
\newblock \url{https://www.synopsys.com/verification/debug/verdi.html}.

\bibitem{srcLoc_2021}
Get source location details from ir code (function pass) 2021.
\newblock
  \url{https://discourse.llvm.org/t/get-source-location-details-from-ir-code-function-pass/57372},
  Jan 2021.

\bibitem{xls}
{XLS: Accelerated HW Synthesis}.
\newblock \url{https://github.com/google/xls}, 2021.
\newblock {Online; accessed on 9 August 2021}.

\bibitem{circt}
{CIRCT: Circuit IR Compilers and Tools}.
\newblock \url{https://github.com/llvm/circt}, 2022.
\newblock {Online; accessed on 12 August 2022}.

\bibitem{Asanović:EECS-2016-17}
Krste Asanović, Rimas Avizienis, Jonathan Bachrach, Scott Beamer, David
  Biancolin, Christopher Celio, Henry Cook, Daniel Dabbelt, John Hauser, Adam
  Izraelevitz, Sagar Karandikar, Ben Keller, Donggyu Kim, John Koenig, Yunsup
  Lee, Eric Love, Martin Maas, Albert Magyar, Howard Mao, Miquel Moreto, Albert
  Ou, David~A. Patterson, Brian Richards, Colin Schmidt, Stephen Twigg, Huy Vo,
  and Andrew Waterman.
\newblock The rocket chip generator.
\newblock Technical Report UCB/EECS-2016-17, Apr 2016.

\bibitem{chisel}
Jonathan Bachrach, Huy Vo, Brian Richards, Yunsup Lee, Andrew Waterman, Rimas
  Avi{\v{z}}ienis, John Wawrzynek, and Krste Asanovi{\'c}.
\newblock Chisel: constructing hardware in a scala embedded language.
\newblock In {\em DAC Design Automation Conference 2012}, pages 1212--1221.
  IEEE, 2012.

\bibitem{8639205}
G.D. Balogh, G.R. Mudalige, I.Z. Reguly, S.F. Antao, and C.~Bertolli.
\newblock Op2-clang: A source-to-source translator using clang/llvm libtooling.
\newblock In {\em 2018 IEEE/ACM 5th Workshop on the LLVM Compiler
  Infrastructure in HPC (LLVM-HPC)}, pages 59--70, 2018.

\bibitem{bayati2013message}
Mohsen Bayati, David~F Gleich, Amin Saberi, and Ying Wang.
\newblock Message-passing algorithms for sparse network alignment.
\newblock {\em ACM Transactions on Knowledge Discovery from Data (TKDD)},
  7(1):1--31, 2013.

\bibitem{chisel2022}
chipsalliance.
\newblock Utils: add source locators by mwachs5 · pull request 2496 ·
  chipsalliance/chisel.
\newblock \url{https://github.com/chipsalliance/chisel3/pull/2496}, Apr.

\bibitem{chisel2016}
chipsalliance.
\newblock Source information via macros, rfc · issue 147 ·
  chipsalliance/chisel.
\newblock \url{https://github.com/chipsalliance/chisel3/issues/147}, Apr 2016.

\bibitem{ghosh2019case}
Prokash Ghosh and Srivastava Rohit.
\newblock Case study: Soc performance verification and static verification of
  rtl parameters.
\newblock In {\em 2019 20th International Workshop on Microprocessor/SoC Test,
  Security and Verification (MTV)}, pages 65--72. IEEE, 2019.

\bibitem{marmot_asic}
{Hagiwara-shc, Jeff DiCorpo, Manar, Marwan Abbas, Kareem Farid, Mohamed Kassem,
  R. Timothy Edwards, Russell Friesenhahn, matt venn, Amr A. Gouhar, Mohamed
  Shalan}.
\newblock {Marmot RISC-V SoC}.
\newblock \url{https://github.com/Hagiwara-shc/marmot_asic.git}, 2022.
\newblock {Online; accessed on March 2024}.

\bibitem{heimann2021refining}
Mark Heimann, Xiyuan Chen, Fatemeh Vahedian, and Danai Koutra.
\newblock Refining network alignment to improve matched neighborhood
  consistency.
\newblock In {\em Proceedings of the 2021 SIAM International Conference on Data
  Mining (SDM)}, pages 172--180. SIAM.

\bibitem{heimann2018regal}
Mark Heimann, Haoming Shen, Tara Safavi, and Danai Koutra.
\newblock Regal: Representation learning-based graph alignment.
\newblock In {\em Proceedings of the 27th ACM international conference on
  information and knowledge management}, pages 117--126, 2018.

\bibitem{opensta}
{James Cherry}.
\newblock {OpenSTA}.
\newblock \url{https://github.com/abk-openroad/OpenSTA}.
\newblock Online; accessed on 5 September 2019.

\bibitem{PipelinedCPU}
jlpteaching.
\newblock dinocpu/cpu.scala (pipelinedcpu) at main · jlpteaching/dinocpu.
\newblock
  \url{https://github.com/jlpteaching/dinocpu/blob/main/src/main/scala/pipelined/cpu.scala}.

\bibitem{SingleCycleCPU}
jlpteaching.
\newblock dinocpu/cpu.scala (singlecyclecpu) at main · jlpteaching/dinocpu.
\newblock
  \url{https://github.com/jlpteaching/dinocpu/blob/main/src/main/scala/single-cycle/cpu.scala}.

\bibitem{kazemi2016network}
Ehsan Kazemi.
\newblock Network alignment: Theory, algorithms, and applications.
\newblock Technical report, EPFL, 2016.

\bibitem{kobler2012graph}
Johannes Kobler, Uwe Sch{\"o}ning, and Jacobo Tor{\'a}n.
\newblock {\em The graph isomorphism problem: its structural complexity}.
\newblock Springer Science \& Business Media, 2012.

\bibitem{v350chisel1}
Jack Koenig.
\newblock Release chisel v3.5.0 · chipsalliance/chisel.
\newblock \url{https://github.com/chipsalliance/chisel3/releases/tag/v3.5.0},
  Jan 2001.

\bibitem{v355chisel2001}
Jack Koenig.
\newblock Release chisel v3.5.5 · chipsalliance/chisel.
\newblock \url{https://github.com/chipsalliance/chisel3/releases/tag/v3.5.5},
  Nov 2001.

\bibitem{vM1}
Jack Koenig.
\newblock Release chisel v3.6.0-m1 · chipsalliance/chisel.
\newblock
  \url{https://github.com/chipsalliance/chisel3/releases/tag/v3.6.0-M1}, Dec
  2001.

\bibitem{vM2}
Jack Koenig.
\newblock Release chisel v3.6.0-m2 · chipsalliance/chisel.
\newblock
  \url{https://github.com/chipsalliance/chisel3/releases/tag/v3.6.0-M2}, Jan
  2001.

\bibitem{liu2016aligning}
Li~Liu, William~K Cheung, Xin Li, and Lejian Liao.
\newblock Aligning users across social networks using network embedding.
\newblock In {\em Ijcai}, volume~16, pages 1774--1780, 2016.

\bibitem{lund2021design}
Ryan Lund.
\newblock {\em Design and Application of a Co-Simulation Framework for Chisel}.
\newblock PhD thesis, MA thesis. EECS Department, University of California,
  Berkeley, 2021.

\bibitem{maidhili2020reset}
Kakarlamudi~Lakshmi Maidhili, Fazal Noorbasha, Allamsetty Vamsi, and
  Kakarla~Hari Kishore.
\newblock Reset logic verification of an iod at system on chip level using
  gatesim.
\newblock {\em International Journal}, 8(7), 2020.

\bibitem{lec_formality_design_reuse}
Priyambada Mishra.
\newblock Understanding logic equivalence check (lec) flow and its challenges,
  and proposed solution.
\newblock
  \url{https://www.design-reuse.com/articles/51622/understanding-logic-equivalence-check-lec-flow-and-its-challenges-and-proposed-solution.html}.
\newblock Accessed: 2024-10-07.

\bibitem{ibtida}
{Muhammad Hadir Khan, R. Timothy Edwards, Amr A. Gouhar, Manar, Mohamed Kassem,
  Mohamed Shalan, AireenAmirJalal, matt venn, Jeff DiCorpo}.
\newblock {SoC - Google SKY130 Shuttle (ibtida)}.
\newblock \url{https://github.com/hadirkhan10/caravel_ibtida_soc.git}, 2021.
\newblock {Online; accessed on March 2024}.

\bibitem{netolicka2005equivalence}
Karol{\'\i}na Netolick{\'a} et~al.
\newblock {\em Equivalence checking of retimed circuits}.
\newblock PhD thesis, Massachusetts Institute of Technology, 2005.

\bibitem{pal2023equality}
Anjali Pal, Brett Saiki, Ryan Tjoa, Cynthia Richey, Amy Zhu, Oliver Flatt, Max
  Willsey, Zachary Tatlock, and Chandrakana Nandi.
\newblock Equality saturation theory exploration {\`a} la carte.
\newblock {\em Proceedings of the ACM on Programming Languages},
  7(OOPSLA2):1034--1062, 2023.

\bibitem{palnitkar2003verilog}
Samir Palnitkar.
\newblock {\em Verilog HDL: a guide to digital design and synthesis}, volume~1.
\newblock Prentice Hall Professional, 2003.

\bibitem{lec_des}
Prathmesh~Oza Pratik~Patel and Rakesh Parmar.
\newblock {A Guide on Logical Equivalence Checking - Flow, Challenges, and
  Benefits}.
\newblock
  \url{https://www.design-reuse.com/articles/45547/a-guide-on-logical-equivalence-checking-flow-challenges-and-benefits.html}.

\bibitem{Shashank_06_mac}
{Shashank, Jeff DiCorpo, Manar, Marwan Abbas, Kareem Farid, Mohamed Kassem, R.
  Timothy Edwards, Russell Friesenhahn, Johan Euphrosine, Amr A. Gouhar, matt
  venn, Mohamed Gaber, Mohamed Shalan}.
\newblock {Dual mac unit array with a single sigmoid activation function}.
\newblock \url{https://github.com/Shashank-06/iit_indore_neuron.git}, 2022.
\newblock {Online; accessed on March 2024}.

\bibitem{smith2024there}
Gus~Henry Smith, Zachary~D Sisco, Thanawat Techaumnuaiwit, Jingtao Xia, Vishal
  Canumalla, Andrew Cheung, Zachary Tatlock, Chandrakana Nandi, and Jonathan
  Balkind.
\newblock There and back again: A netlist's tale with much egraphin'.
\newblock {\em arXiv preprint arXiv:2404.00786}, 2024.

\bibitem{synopsys:design-compiler}
{Synopsys Inc.}
\newblock {Design Compiler User Guide}.

\bibitem{synopsys:primetime}
{Synopsys, Inc.}
\newblock Primetime static timing analysis.
\newblock
  \url{https://www.synopsys.com/implementation-and-signoff/signoff/primetime.html}.
\newblock Online; accessed on 26 April 2019.

\bibitem{thomas2008verilog}
Donald Thomas and Philip Moorby.
\newblock {\em The Verilog{\textregistered} hardware description language}.
\newblock Springer Science \& Business Media, 2008.

\bibitem{livehd}
Sheng-Hong Wang, Rafael~Trapani Possignolo, Haven~Blake Skinner, and Jose
  Renau.
\newblock {LiveHD: A Productive Live Hardware Development Flow}.
\newblock {\em IEEE Micro}, 40(4):67--75, 2020.

\bibitem{wolf2021yosys}
Clifford Wolf.
\newblock Yosys manual.
\newblock {\em Retrieved January}, 16:2021, 2021.

\bibitem{yosys}
Clifford Wolf.
\newblock {Yosys Open SYnthesis Suite}.
\newblock \url{https://github.com/YosysHQ/yosys}, 2022.
\newblock {Online; accessed on December 2022}.

\bibitem{zemlyachenko1985graph}
Viktor~N Zemlyachenko, Nickolay~M Korneenko, and Regina~I Tyshkevich.
\newblock Graph isomorphism problem.
\newblock {\em Journal of Soviet Mathematics}, 29:1426--1481, 1985.

\bibitem{zhang2016final}
Si~Zhang and Hanghang Tong.
\newblock Final: Fast attributed network alignment.
\newblock In {\em Proceedings of the 22nd ACM SIGKDD international conference
  on knowledge discovery and data mining}, pages 1345--1354, 2016.

\bibitem{zhu2022caper}
Jing Zhu, Danai Koutra, and Mark Heimann.
\newblock Caper: Coarsen, align, project, refine-a general multilevel framework
  for network alignment.
\newblock In {\em Proceedings of the 31st ACM International Conference on
  Information \& Knowledge Management}, pages 4747--4751, 2022.

\end{thebibliography}

\end{document}